\documentstyle[prd,aps,preprint,tighten,epsfig]{revtex}
\begin{document}

\draft

\title{Model-independent access to the structure of quark flavor mixing}
\author{{\bf Zhi-zhong Xing} \thanks{E-mail:
xingzz@ihep.ac.cn}}
\address{Institute of High Energy Physics and Theoretical Physics Center
for Science Facilities, \\
Chinese Academy of Sciences, Beijing 100049, China \\
Center for High Energy Physics, Peking University, Beijing 100871,
China} \maketitle

\begin{abstract}
We show that the observed pattern of quark flavor mixing, such as
$|V^{}_{us}| \simeq |V^{}_{cd}|$, $|V^{}_{cb}| \simeq |V^{}_{ts}|$,
$|V^{}_{cd}/V^{}_{td}| \simeq |V^{}_{cs}/V^{}_{ts}| \simeq
|V^{}_{tb}/V^{}_{cb}|$ and $|V^{}_{ub}/V^{}_{cb}| <
|V^{}_{td}/V^{}_{ts}|$, can essentially be understood in the chiral
and heavy quark mass limits. In particular, the phenomenologically
successful relations $|V^{}_{ub}/V^{}_{cb}| \sim
\sqrt{m^{}_u/m^{}_c}$ and $|V^{}_{td}/V^{}_{ts}| \simeq
\sqrt{m^{}_d/m^{}_s}$ can be reasonably conjectured in the $m^{}_b
\to \infty$ and $m^{}_t \to \infty$ limits, respectively. We stress
that the strength of weak CP violation in the quark sector is
determined by the moduli of the four corner elements of the
Cabibbo-Kobayashi-Maskawa matrix. A comparison between strong and
weak CP-violating effects in the standard model is also made.
\end{abstract}


\newpage

\section{Introduction}

The standard model (SM) has proved to be very successful in
describing the fundamental properties of elementary particles and
their interactions, but it has to be extended in the flavor sector
by covering the fact that the three known neutrinos are not massless
and different lepton flavors can mix \cite{PDG}. In this case even a
minimal extension of the SM involves twenty (or twenty-two)
flavor parameters, among which twelve are masses, six are flavor
mixing angles and two (or four) are CP-violating phases if the
massive neutrinos are of the Dirac (or Majorana) nature
\footnote{In this regard we have assumed the $3\times 3$ lepton
flavor mixing matrix to be unitary, regardless of the origin of tiny
neutrino masses. Here the effective strong CP-violating parameter
$\overline{\theta}$ is not taken into account, but it will be
briefly discussed in section IV.}.
Determining the values of these parameters and understanding why
they are what they are constitute a central part of today's particle
physics. However, we have to confront some flavor puzzles revealed
by current experimental data.

The flavor puzzles include why up-type quarks, down-type quarks and
charged leptons all have strong mass hierarchies (i.e., $m^{}_u \ll
m^{}_c \ll m^{}_t$, $m^{}_d \ll m^{}_s \ll m^{}_b$ and $m^{}_e \ll
m^{}_\mu \ll m^{}_\tau$) at a given energy scale; why the masses of
three neutrinos are extremely small in comparison with those of nine
charged fermions; why the six off-diagonal elements of the
Cabibbo-Kobayashi-Maskawa (CKM) quark flavor mixing matrix $V$
\cite{CKM} are strongly suppressed such that the three mixing angles
are very small; why the Maki-Nakagawa-Sakata-Pontecorvo (MNSP)
lepton flavor mixing matrix $U$ \cite{MNSP} contains two relatively
large mixing angles and the third one is not strongly suppressed
either; why the patterns of $V$ and $U$ are so different but their
smallest matrix elements are both located at the upper-right corner
(i.e., $V^{}_{ub}$ and $U^{}_{e 3}$) \cite{X2012}; how the origin of
CP violation is correlated with the origin of fermion masses; and so
on. In the lack of a complete flavor theory capable of predicting
the flavor structures of leptons and quarks or revealing possible
symmetries behind them, it is a big challenge to answer even a part
of the aforecited questions. The great ideas like grand
unifications, supersymmetries and extra dimensions are still not
very helpful to solve the flavor puzzles, and the exercises of
various group languages or flavor symmetries turn out to be too
divergent to converge to something unique \cite{SV}.

In this paper we shall follow a purely phenomenological way to
speculate whether the observed pattern of quark flavor mixing can be
partly understood in some reasonable limits of quark masses. This
starting point of view is more or less motivated by two useful
working symmetries in understanding the strong interactions of
quarks and hadrons by means of the quantum chromodynamics (QCD) or
an effective field theory based on the QCD \cite{W}: the chiral
quark symmetry (i.e., $m^{}_u, m^{}_d, m^{}_s \to 0$) and the heavy
quark symmetry (i.e., $m^{}_c, m^{}_b, m^{}_t \to \infty$). The
reason for the usefulness of these two symmetries is simply that the
masses of the light quarks are far below the typical QCD scale
$\Lambda^{}_{\rm QCD} \sim 0.2$ GeV, whereas the masses of the heavy
quarks are far above it. Because the elements of the CKM matrix $V$
are dimensionless and their magnitudes lie in the range of 0 to 1,
they are in general expected to depend on the mass ratios of the
lighter quarks to the heavier quarks. The mass limits corresponding
to the chiral and heavy quark symmetries are therefore equivalent to
setting the relevant mass ratios to zero, and they are possible to
help reveal a part of the salient features of $V$. In this spirit,
some preliminary attempts have been made to look at the quark flavor
mixing pattern in the $m^{}_u, m^{}_d \to 0$ or $m^{}_t, m^{}_b
\to\infty$ limits \cite{F87}.

The present work aims to show that it is actually possible to gain
an insight into the observed pattern of quark flavor mixing in the
chiral and heavy quark mass limits. Such a model-independent access
to the underlying quark flavor structure can at least explain why
$|V^{}_{us}| \simeq |V^{}_{cd}|$ and $|V^{}_{cb}| \simeq
|V^{}_{ts}|$ hold to a good degree of accuracy, why
$|V^{}_{cd}/V^{}_{td}| \simeq |V^{}_{cs}/V^{}_{ts}| \simeq
|V^{}_{tb}/V^{}_{cb}|$ is a reasonable approximation, and why
$|V^{}_{ub}/V^{}_{cb}|$ should be smaller than
$|V^{}_{td}/V^{}_{ts}|$. In particular, the phenomenologically
successful relations $|V^{}_{ub}/V^{}_{cb}| \sim
\sqrt{m^{}_u/m^{}_c}$ and $|V^{}_{td}/V^{}_{ts}| \simeq
\sqrt{m^{}_d/m^{}_s}$ can be reasonably conjectured in the heavy
quark mass limits. We also point out that the strength of CP
violation in the quark sector is simply determined by the product of
the four corner elements of $V$ (i.e., $|V^{}_{ud}|$, $|V^{}_{ub}|$,
$|V^{}_{td}|$ and $|V^{}_{tb}|$), based on the experimental fact
that two of the CKM unitarity triangles are almost the right
triangles. This interesting observation motivates us to pay more
attention to a particular parametrization of $V$ \cite{GX}, in which
the three flavor mixing angles are all comparable with the Cabibbo
angle and the tiny CP-violating phase only show up in the four
corners of $V$. Finally, we make a brief comment on the effect of
strong CP violation and then compare its strength with that of weak
CP violation.

\section{The CKM matrix in the quark mass limits}

We begin with the weak charged-current interactions of six quarks in
their mass basis:
\begin{eqnarray}
-{\cal L}^{}_{\rm cc} = \frac{g}{\sqrt{2}} \ \overline{\left(u ~~ c
~~ t\right)^{}_{\rm L}} \ \gamma^\mu \ V \left(\matrix{d \cr s \cr
b\cr}\right)^{}_{\rm L} W^+_\mu + {\rm h.c.} \; ,
\end{eqnarray}
where the CKM matrix $V$ measures a nontrivial mismatch between the
flavor and mass eigenstates and can be decomposed into $V =
O^\dagger_{\rm u} O^{}_{\rm d}$ with $O^{}_{\rm u}$ and $O^{}_{\rm
d}$ being the unitary transformations responsible for the
diagonalizations of the up- and down-type quark mass matrices in the
flavor basis. Namely,
\begin{eqnarray}
O^\dagger_{\rm u} H^{}_{\rm u} O^{}_{\rm u} & = &
O^\dagger_{\rm u} M^{}_{\rm u} M^\dagger_{\rm u}
O^{}_{\rm u} = {\rm Diag} \left\{m^2_u , m^2_c , m^2_t \right\} \; ,
\nonumber \\
O^\dagger_{\rm d} H^{}_{\rm d} O^{}_{\rm d} & = &
O^\dagger_{\rm d} M^{}_{\rm d} M^\dagger_{\rm d}
O^{}_{\rm d} = {\rm Diag} \left\{m^2_d , m^2_s , m^2_b \right\} \; ,
\end{eqnarray}
where $H^{}_{\rm u}$ and $H^{}_{\rm d}$ are defined to be Hermitian.
To be explicit, the nine elements of $V$ read
\begin{eqnarray}
V^{}_{\alpha i} & = & \sum^3_{k=1}
(O^{}_{\rm u})^*_{k \alpha} (O^{}_{\rm d})^{}_{k i} \; , ~~
\end{eqnarray}
where $\alpha$ and $i$ run over $(u, c, t)$ and $(d, s, b)$,
respectively. Because of Eqs. (2) and (3), the dimensionless
$V^{}_{\alpha i}$ elements are expected to be more or less dependent
on the quark mass ratios and nontrivial phase differences between
$M^{}_{\rm u}$ and $M^{}_{\rm d}$. Given current experimental data
and the unitarity of $V$, the magnitudes of all the nine CKM matrix
elements have been determined to an impressively good degree of
accuracy \cite{PDG}:
\begin{eqnarray}
|V| = \left(\matrix{ 0.97427 \pm 0.00015 & 0.22534 \pm 0.00065 &
0.00351^{+0.00015}_{-0.00014} \cr 0.22520 \pm 0.00065 & 0.97344 \pm
0.00016 & 0.0412^{+0.0011}_{-0.0005} \cr
0.00867^{+0.00029}_{-0.00031} & 0.0404^{+0.0011}_{-0.0005} &
0.999146^{+0.000021}_{-0.000046} \cr} \right) \; .
\end{eqnarray}
We shall show that the strong mass hierarchies of up- and down-type
quarks allow us to account for a part of the observed flavor mixing
properties in a model-independent way.

\subsection{$H^{}_{\rm u}$ and $H^{}_{\rm d}$ in the quark mass limits}

In general, the mass limit $m^{}_u \to 0$ (or $m^{}_d \to 0$) does
not correspond to a unique form of $H^{}_{\rm u}$ (or $H^{}_{\rm
d}$). The reason is simply that the form of a quark mass matrix is
always basis-dependent. Without loss of generality, one may choose a
particular flavor basis such that $H^{}_{\rm u}$ and $H^{}_{\rm d}$
can be written as
\begin{eqnarray}
\lim_{m^{}_u \to 0} H^{}_{\rm u} & = & \left( \matrix{0 & 0 & 0 \cr
0 & ~ \times ~ & \times \cr 0 &
\times & \times \cr} \right) \; ,
\nonumber \\
\lim_{m^{}_d \to 0} H^{}_{\rm d} & = & \left( \matrix{0 & 0 & 0 \cr
0 & ~ \times ~ & \times \cr 0 &
\times & \times \cr} \right) \; ,
\end{eqnarray}
in which ``$\times$" denotes an arbitrary  nonzero element. To prove
that Eq. (5) is the result of a basis choice instead of an
assumption, we refer the reader to Appendix A.

When the mass of a given quark goes to infinity, we argue that it
becomes decoupled from the masses of other quarks. In this sense we
may choose a specific flavor basis where $H^{}_{\rm u}$ and
$H^{}_{\rm d}$ can be written as
\begin{eqnarray}
\lim_{m^{}_t \to \infty} H^{}_{\rm u} & = &
\left( \matrix{\times & ~ \times ~ & 0 \cr \times &
\times & 0 \cr 0 & 0 & \infty \cr} \right) \; ,
\nonumber \\
\lim_{m^{}_b \to \infty} H^{}_{\rm d} & = &
\left( \matrix{\times & ~ \times ~ & 0 \cr \times &
\times & 0 \cr 0 & 0 & \infty \cr} \right) \; .
\end{eqnarray}
In other words, the $3\times 3$ Hermitian matrices $H^{}_{\rm u}$
and $H^{}_{\rm d}$ can be simplified to the effective $2\times 2$
Hermitian matrices in either the chiral quark mass limit or the
heavy quark mass limit. In view of the fact that $m^{}_u \ll m^{}_c
\ll m^{}_t$ and $m^{}_d \ll m^{}_s \ll m^{}_b$ hold at an arbitrary
energy scale \cite{XZZ}, we believe that Eqs. (5) and (6) are
phenomenologically reasonable and can help explain some of the
observed properties of quark flavor mixing in a model-independent
way. Let us go into details in the following.

\subsection{Why $|V^{}_{us}| \simeq |V^{}_{cd}|$ and $|V^{}_{cb}|
\simeq |V^{}_{ts}|$ hold?}

A glance at Eq. (4) tells us that $|V^{}_{us}| \simeq |V^{}_{cd}|$
is an excellent approximation. Such an approximate equality can be
well understood in the heavy quark mass limits, where Hermitian
$H^{}_{\rm u}$ and $H^{}_{\rm d}$ may take the form of Eq. (6). In
this case the unitary matrices $O^{}_{\rm u}$ and $O^{}_{\rm d}$
used to diagonalize $H^{}_{\rm u}$ and $H^{}_{\rm d}$ in Eq. (2) can
be expressed as
\begin{eqnarray}
\lim_{m^{}_t \to \infty} O^{}_{\rm u} & = &
P^{}_{12} \left( \matrix{c^{}_{12} & s^{}_{12} & 0 \cr
-s^{}_{12} & c^{}_{12} & 0 \cr 0 & 0 & 1 \cr} \right) \; ,
\nonumber \\
\lim_{m^{}_b \to \infty} O^{}_{\rm d} & = &
P^\prime_{12} \left( \matrix{c^{\prime}_{12} & s^{\prime}_{12} & 0
\cr -s^{\prime}_{12} & c^{\prime}_{12} & 0 \cr 0 & 0 & 1 \cr}
\right) \; ,
\end{eqnarray}
where $c^{(\prime)}_{12} \equiv \cos\vartheta^{(\prime)}_{12}$,
$s^{(\prime)}_{12} \equiv \sin\vartheta^{(\prime)}_{12}$, and
$P^{(\prime)}_{12} = {\rm Diag}\left\{e^{i\phi^{(\prime)}_{12}}, 1,
1 \right\}$. So Eq. (7) yields
\begin{eqnarray}
\left|V^{}_{us}\right| = \left|c^{}_{12} s^\prime_{12} -
s^{}_{12} c^\prime_{12} e^{i\Delta^{}_{12}}\right| =
\left|V^{}_{cd}\right|
\end{eqnarray}
in the $m^{}_t \to \infty$ and $m^{}_b \to \infty$ limits, where
$\Delta^{}_{12} \equiv \phi^\prime_{12} - \phi^{}_{12}$ denotes
the nontrivial phase difference between the up- and down-quark
sectors. Since
$m^{}_u/m^{}_c \sim m^{}_c/m^{}_t \sim \lambda^4$ and $m^{}_d/m^{}_s
\sim m^{}_s/m^{}_b \sim \lambda^2$ hold \cite{XZZ}, where $\lambda
\equiv \sin\theta^{}_{\rm C} \simeq 0.22$ with $\theta^{}_{\rm C}$
being the Cabibbo angle, the mass limits taken above are apparently
a good approximation. Therefore, we conclude that the approximate
equality $|V^{}_{us}| \simeq |V^{}_{cd}|$ is attributed to the fact
that $m^{}_t \gg m^{}_u, m^{}_c$ and $m^{}_b \gg m^{}_d, m^{}_s$
hold
\footnote{Quantitatively, $|V^{}_{us}| \simeq |V^{}_{cd}| \simeq
\lambda$ holds. Hence $s^{}_{12} \simeq \sqrt{m^{}_u/m^{}_c} \simeq
\lambda^2$ and $s^\prime_{12} \simeq \sqrt{m^{}_d/m^{}_s} \simeq
\lambda$ are often conjectured and can easily be derived from some
ans$\rm\ddot{a}$tze of quark mass matrices \cite{F77}.}.

One may similarly consider the chiral quark mass limits $m^{}_u \to
0$ and $m^{}_d \to 0$ in order to understand why $|V^{}_{ts}| \simeq
|V^{}_{cb}|$ holds. In this case, Eq. (5) leads us to
\begin{eqnarray}
\lim_{m^{}_u \to 0} O^{}_{\rm u} & = &
P^{}_{23} \left( \matrix{1 & 0 & 0 \cr 0 & c^{}_{23} &
s^{}_{23} \cr 0 & -s^{}_{23} & c^{}_{23} \cr} \right) \; ,
\nonumber \\
\lim_{m^{}_d \to 0} O^{}_{\rm d} & = &
P^\prime_{23} \left( \matrix{1 & 0 & 0 \cr 0 & c^{\prime}_{23} &
s^{\prime}_{23} \cr 0 & -s^{\prime}_{23} & c^{\prime}_{23} \cr}
\right) \; ,
\end{eqnarray}
where $c^{(\prime)}_{23} \equiv \cos\vartheta^{(\prime)}_{23}$,
$s^{(\prime)}_{23} \equiv \sin\vartheta^{(\prime)}_{23}$, and
$P^{(\prime)}_{23} = {\rm Diag}\left\{1, 1,
e^{i\phi^{(\prime)}_{23}}\right\}$. We are therefore left with
\begin{eqnarray}
\left|V^{}_{cb}\right| = \left|c^{}_{23} s^\prime_{23} - s^{}_{23}
c^\prime_{23} e^{i\Delta^{}_{23}} \right| = \left|V^{}_{ts}\right|
\end{eqnarray}
in the $m^{}_u \to 0$ and $m^{}_d \to 0$ limits, where
$\Delta^{}_{23} \equiv \phi^\prime_{23} - \phi^{}_{23}$ stands for
the nontrivial phase difference between the up- and down-quark
sectors. This model-independent result is also in good agreement
with the experimental data $|V^{}_{cb}| \simeq |V^{}_{ts}|$ as given
in Eq. (4). In other words, the approximate equality $|V^{}_{cb}|
\simeq |V^{}_{ts}|$ is a natural consequence of $m^{}_u \ll m^{}_c,
m^{}_t$ and $m^{}_d \ll m^{}_s, m^{}_b$ in no need of any specific
assumptions
\footnote{It is possible to obtain the quantitative relationship
$|V^{}_{cb}| \simeq |V^{}_{ts}| \simeq \lambda^2$ through $s^{}_{23}
\simeq m^{}_c/m^{}_t \simeq \lambda^4$ and $s^\prime_{23} \simeq
m^{}_s/m^{}_b \simeq \lambda^2$ from a number of ans$\rm\ddot{a}$tze
of quark mass matrices \cite{FX00}.}.

\subsection{Why $|V^{}_{cd}/V^{}_{td}| \simeq |V^{}_{cs}/V^{}_{ts}|
 \simeq |V^{}_{tb}/V^{}_{cb}|$ holds?}

Given the magnitudes of the CKM matrix elements in Eq. (4),  it is
straightforward to obtain $|V^{}_{cd}/V^{}_{td}| \simeq 26.0$,
$|V^{}_{cs}/V^{}_{ts}| \simeq 24.1$ and $|V^{}_{tb}/V^{}_{cb}|
\simeq 24.3$. These numbers imply $|V^{}_{cd}/V^{}_{td}| \simeq
|V^{}_{cs}/V^{}_{ts}| \simeq |V^{}_{tb}/V^{}_{cb}|$ as a reasonably
good approximation, which has not come into notice in the
literature. We find that such an approximate relation becomes exact
in the mass limits $m^{}_u \to 0$ and $m^{}_b \to \infty$. To be
explicit,
\begin{eqnarray}
V = \lim_{m^{}_u \to 0} O^{\dagger}_{\rm u} \lim_{m^{}_b \to \infty}
O^{}_{\rm d} = P^\prime_{12}
\left(\matrix{c^\prime_{12} & s^\prime_{12} & 0 \cr
-c^{}_{23} s^\prime_{12} & c^{}_{23} c^\prime_{12} & -s^{}_{23} \cr
-s^{}_{23} s^\prime_{12} & s^{}_{23} c^\prime_{12} & c^{}_{23} \cr}
\right) P^\dagger_{23} \; ,
\end{eqnarray}
where Eqs. (7) and (9) have been used. Therefore,
\begin{eqnarray}
\left|\frac{V^{}_{cd}}{V^{}_{td}}\right| =
\left|\frac{V^{}_{cs}}{V^{}_{ts}}\right| =
\left|\frac{V^{}_{tb}}{V^{}_{cb}}\right| =
\left|\cot\vartheta^{}_{23}\right| \;
\end{eqnarray}
holds in the chosen quark mass limits, which assure the smallest CKM
matrix element $V^{}_{ub}$ to vanish. This simple result is
essentially consistent with the experimental data if
$\vartheta^{}_{23} \simeq 2.35^\circ$ is taken
\footnote{This numerical estimate implies $\tan\vartheta^{}_{23}
\simeq \lambda^2 \simeq \sqrt{m^{}_c/m^{}_t}~$, which can easily be
derived from the Fritzsch ansatz of quark mass matrices \cite{F78}.}.
Note that the quark mass limits $m^{}_t \to \infty$ and $m^{}_d \to
0$ are less favored because they predict both $|V^{}_{td}| =0$ and
$|V^{}_{us}/V^{}_{ub}| = |V^{}_{cs}/V^{}_{cb}| =
|V^{}_{tb}/V^{}_{ts}|$, which are in conflict with current
experimental data. In particular, the limit $|V^{}_{ub}| =0$ is
apparently closer to reality than the limit $|V^{}_{td}| =0$.

But why $V^{}_{ub}$ is smaller in magnitude than all the other CKM
matrix elements remains a puzzle, since it is difficult for us to
judge that the quark mass limits $m^{}_u \to \infty$ and $m^{}_b \to
0$ should make more sense than the quark mass limits $m^{}_t \to
\infty$ and $m^{}_d \to 0$ from a phenomenological point of view.
The experimental data in Eq. (4) indicate $|V^{}_{td}| \gtrsim
2|V^{}_{ub}|$ and $|V^{}_{ts}| \simeq |V^{}_{cb}|$. So a comparison
between the ratios $|V^{}_{ub}/V^{}_{cb}|$ and
$|V^{}_{td}/V^{}_{ts}|$ might be able to tell us an acceptable
reason for $|V^{}_{td}| > |V^{}_{ub}|$.

\subsection{Why $|V^{}_{ub}/V^{}_{cb}|$ is smaller than
$|V^{}_{td}/V^{}_{ts}|$?}

With the help of Eqs. (3), (6) and (7), we can calculate the ratios
$|V^{}_{ub}/V^{}_{cb}|$ and $|V^{}_{td}/V^{}_{ts}|$ in the
respective heavy quark mass limits:
\begin{eqnarray}
\lim_{m^{}_b \to \infty} \left|\frac{V^{}_{ub}}{V^{}_{cb}}\right| &
= & \left|\frac{(O^{}_{\rm u})^{}_{3 u}}{(O^{}_{\rm u})^{}_{3
c}}\right| \; ,
\nonumber \\
\lim_{m^{}_t \to \infty} \left|\frac{V^{}_{td}}{V^{}_{ts}}\right| &
= & \left|\frac{(O^{}_{\rm d})^{}_{3 d}}{(O^{}_{\rm d})^{}_{3
s}}\right| \; . ~~
\end{eqnarray}
This result is quite nontrivial in the sense that
$|V^{}_{ub}/V^{}_{cb}|$ turns out to be independent of the mass
ratios of three down-type quarks in the $m^{}_b \to \infty$ limit,
and $|V^{}_{td}/V^{}_{ts}|$ has nothing to do with the mass ratios
of three up-type quarks in the $m^{}_t \to \infty$ limit. In
particular, the flavor indices showing up on the right-hand side of
Eq. (13) is rather suggestive: $|V^{}_{ub}/V^{}_{cb}|$ is relevant
to $u$ and $c$ quarks, and $|V^{}_{td}/V^{}_{ts}|$ depends on $d$
and $s$ quarks. We are therefore encouraged to conjecture that
$|V^{}_{ub}/V^{}_{cb}|$ (or $|V^{}_{td}/V^{}_{ts}|$) should be a
simple function of the mass ratio $m^{}_u/m^{}_c$ (or
$m^{}_d/m^{}_s$) in the $m^{}_t \to \infty$ (or $m^{}_b \to \infty$)
limit.

Given the renormalized quark mass values $m^{}_u =
1.38^{+0.42}_{-0.41}$ MeV, $m^{}_d = 2.82 \pm 0.048$ MeV, $m^{}_s =
57^{+18}_{-12}$ MeV and $m^{}_c = 0.638^{+0.043}_{-0.084}$ GeV at
the energy scale $\mu = M^{}_Z$ \cite{XZZ}, the simplest
phenomenological conjectures turn out to be
\begin{eqnarray}
\lim_{m^{}_b \to \infty} \left|\frac{V^{}_{ub}}{V^{}_{cb}}\right| &
\simeq & c^{}_1 \sqrt{\frac{m^{}_u}{m^{}_c}} \; ,
\nonumber \\
\lim_{m^{}_t \to \infty} \left|\frac{V^{}_{td}}{V^{}_{ts}}\right| &
\simeq & c^{}_2 \sqrt{\frac{m^{}_d}{m^{}_s}} \; ,
\end{eqnarray}
where $c^{}_1$ and $c^{}_2$ are the ${\cal O}(1)$ coefficients. In
view of $\sqrt{m^{}_u/m^{}_c} \simeq \lambda^2$ and
$\sqrt{m^{}_d/m^{}_s} \simeq \lambda$, we expect that
$|V^{}_{ub}/V^{}_{cb}|$ is naturally smaller than
$|V^{}_{td}/V^{}_{ts}|$ in the heavy quark mass limits. Taking
$c^{}_1 =2$ and $c^{}_2 =1$ for example, we obtain
$|V^{}_{ub}/V^{}_{cb}| \simeq 0.093$ and $|V^{}_{td}/V^{}_{ts}|
\simeq 0.222$ from Eq. (14), consistent with the experimental
results $|V^{}_{ub}/V^{}_{cb}| \simeq 0.085$ and
$|V^{}_{td}/V^{}_{ts}| \simeq 0.214$ as given in Eq. (4) \cite{PDG}.
Because $m^{}_t = 172.1 \pm 1.2$ GeV and $m^{}_b =
4.19^{+0.18}_{-0.16}$ GeV at $\mu = M^{}_Z$ \cite{XZZ}, one may
argue that $m^{}_t \to \infty$ is a much better limit and thus the
relation $|V^{}_{td}/V^{}_{ts}| \simeq \sqrt{m^{}_d/m^{}_s}$ has a
good chance to be true. In comparison, $|V^{}_{ub}/V^{}_{cb}| \simeq
2 \sqrt{m^{}_u/m^{}_c}$ suffers from much bigger uncertainties
associated with the values of $m^{}_u$ and $m^{}_c$, and even its
coefficient ``$2$" is questionable.

It is well known that the Fritzsch ansatz of quark mass matrices
\cite{F78} predicts $c^{}_1 \simeq c^{}_2 \simeq 1$. A
straightforward extension of the Fritzsch texture \cite{Du},
\begin{eqnarray}
M^{}_{\rm q} = \left(\matrix{0 & C^{}_{\rm q} & 0 \cr C^*_{\rm q} &
\tilde{B}^{}_{\rm q} & B^{}_{\rm q} \cr 0 & B^*_{\rm q} & A^{}_{\rm
q} \cr} \right)
\end{eqnarray}
with $|A^{}_{\rm q}| \gg |B^{}_{\rm q}| \sim |\tilde{B}^{}_{\rm q}|
\gg |C^{}_{\rm q}|$ (for $\rm q = u$ or $\rm d$), can also lead us
to $c^{}_1 \simeq c^{}_2 \simeq 1$. However, it is always possible
to get $c^{}_1 \simeq \sqrt{2}$ (or $\sqrt{3}$, $2$, $\cdots$)
together with $c^{}_2 \simeq 1$ if the matrix elements $A^{}_{\rm
q}$ and $B^{}_{\rm q}$ have a quite weak hierarchy in magnitude
\cite{FX03}. This observation is interesting, as it implies that the
phenomenological conjectures made in Eq. (14) can be a good starting
point of view for model building in order to understand a possible
correlation between the quark mass spectrum and the flavor mixing
structure.

\section{Implications of the right unitarity triangles}

A rephasing-invariant description of CP violation in the quark
sector is the Jarlskog parameter ${\cal J}^{}_{\rm q}$ defined
through \cite{J}:
\begin{eqnarray}
{\rm Im}\left( V^{}_{\alpha i} V^{}_{\beta j} V^*_{\alpha j}
V^*_{\beta i} \right) = {\cal J}^{}_{\rm q} \sum_{\gamma}
\epsilon^{}_{\alpha\beta\gamma} \sum_{k} \epsilon^{}_{ijk} \; ,
\end{eqnarray}
in which the Greek and Latin subscripts run over $(u,c,t)$ and
$(d,s,b)$, respectively. The unitarity of $V$ leads us to six
triangles in the complex plane, whose areas are all equal to ${\cal
J}^{}_{\rm q}/2$ \cite{FX00}. Among the six CKM unitarity triangles,
$\triangle^{}_s$ and $\triangle^{}_c$ are defined respectively by
the orthogonality relations
\begin{eqnarray}
\triangle^{}_s &:& ~~ V^{}_{ud} V^*_{ub} + V^{}_{cd} V^*_{cb} +
V^{}_{td} V^*_{tb} = 0 \; , \nonumber \\
\triangle^{}_c &:& ~~ V^{}_{tb} V^*_{ub} + V^{}_{ts} V^*_{us} +
V^{}_{td} V^*_{ud} = 0 \; ,
\end{eqnarray}
as illustrated in FIG. 1. They are especially interesting in the sense
that they are essentially the right triangles with a common inner angle
\begin{eqnarray}
\alpha \equiv \arg \left(-\frac{V^{}_{td} V^*_{tb}}{V^{}_{ud}
V^*_{ub}} \right) = \left(89.0^{+4.4}_{-4.2}\right)^\circ \; ,
\end{eqnarray}
as determined by current experimental data \cite{PDG}. Given $\alpha
= 90^\circ$ exactly, it turns out that ${\rm Re} (V^{}_{tb}
V^{}_{ud} V^*_{td} V^*_{ub}) = 0$ holds and thus the
rephasing-invariant quartet $V^{}_{tb} V^{}_{ud} V^*_{td} V^*_{ub}$
is purely imaginary \cite{X09}. In this case $\triangle^{}_s$ and
$\triangle^{}_c$ can be rescaled in such a way that they share a
common side which is equal to
\footnote{One may also obtain ${\cal J}^{}_{\rm q} = |V^{}_{ud}
V^{}_{ub}| \sqrt{|V^{}_{cd} V^{}_{cb}|^2 - |V^{}_{ud} V^{}_{ub}|^2}$
or ${\cal J}^{}_{\rm q} = |V^{}_{ub} V^{}_{tb}| \sqrt{|V^{}_{us}
V^{}_{ts}|^2 - |V^{}_{ub} V^{}_{tb}|^2}$ from FIG. 1 by means of the
Pythagorean theorem. In both cases ${\cal J}^{}_{\rm q}$ is
proportional to the smallest CKM matrix element $|V^{}_{ub}|$. Hence
$|V^{}_{ub}| \neq 0$ is a necessary condition of CP violation.}.
\begin{eqnarray}
{\cal J}^{}_{\rm q} = |V^{}_{ud}| \cdot |V^{}_{ub}| \cdot
|V^{}_{td}| \cdot |V^{}_{tb}| \; ,
\end{eqnarray}
as shown in FIG. 2, where $\triangle^\prime_s$ and
$\triangle^\prime_c$ are the rescaled versions of $\triangle^{}_s$
and $\triangle^{}_c$ with $\alpha = 90^\circ$ (i.e.,
$\triangle^\prime_s: |V^{}_{ud} V^*_{ub}|^2 + V^{}_{ud} V^{}_{cb}
V^*_{ub} V^*_{cd} + i {\cal J}^{}_{\rm q} = 0$ and
$\triangle^\prime_c: |V^{}_{tb} V^*_{ub}|^2 + V^{}_{us} V^{}_{tb}
V^*_{ub} V^*_{ts} + i {\cal J}^{}_{\rm q} = 0$). This result is
quite suggestive: the strength of weak CP violation in the quark
sector is simply determined by the moduli of the four matrix
elements at the four corners of $V$. Note that $|V^{}_{tb}| >
|V^{}_{ud}| > |V^{}_{cs}| \gg |V^{}_{us}| > |V^{}_{cd}| \gg
|V^{}_{cb}| > |V^{}_{ts}| \gg |V^{}_{td}| > |V^{}_{ub}|$ holds for
the nine CKM matrix elements \cite{X96}, so ${\cal J}^{}_{\rm q}$ is
actually equal to the product of two largest and two smallest
elements of $V$ in magnitude. Because $|V^{}_{ud}| \to 1$ and
$|V^{}_{ub}| \to 0$ (or $|V^{}_{tb}| \to 1$ and $|V^{}_{td}| \to 0$)
can be achieved in the $m^{}_u \to 0$ and $m^{}_d \to 0$ limits (or
in the $m^{}_t \to \infty$ and $m^{}_b \to \infty$ limits), we have
${\cal J}^{}_{\rm q} \to 0$ in both case. With the help of Eqs. (4)
and (19), we directly arrive at ${\cal J}^{}_{\rm q} \simeq 2.96
\times 10^{-5}$, which is perfectly consistent with the result
${\cal J}^{}_{\rm q} = \left(2.96^{+0.20}_{-0.16}\right) \times
10^{-5}$ obtained by the Particle Data Group \cite{PDG}.

Given weak CP violation in the quark sector as a corner effect of
the CKM matrix $V$, we have a good reason to consider the following
parametrization of $V$ \cite{GX}:
\begin{eqnarray}
V & = & \left( \matrix{c^{}_y & 0 & s^{}_y \cr 0 & 1 & 0 \cr -s^{}_y
& 0 & c^{}_y \cr} \right) \left( \matrix{ c^{}_x & s^{}_x & 0 \cr
-s^{}_x & c^{}_x & 0 \cr 0 & 0 & e^{-i\delta^{}_{\rm q}} \cr}
\right) \left( \matrix{c^{}_z & 0 & -s^{}_z \cr 0 & 1 & 0 \cr s^{}_z
& 0 & c^{}_z \cr} \right)
\nonumber \\
& = & \left( \matrix{c^{}_x c^{}_y c^{}_z + s^{}_y s^{}_z
e^{-i\delta^{}_{\rm q}} & s^{}_x c^{}_y & -c^{}_x c^{}_y s^{}_z +
s^{}_y c^{}_z e^{-i\delta^{}_{\rm q}} \cr -s^{}_x c^{}_z & c^{}_x &
s^{}_x s^{}_z \cr -c^{}_x s^{}_y c^{}_z + c^{}_y s^{}_z
e^{-i\delta^{}_{\rm q}} & -s^{}_x s^{}_y & c^{}_x s^{}_y s^{}_z +
c^{}_y c^{}_z e^{-i\delta^{}_{\rm q}} \cr} \right) \; ,
\end{eqnarray}
where $c^{}_x \equiv \cos\vartheta^{}_x$ and $s^{}_x \equiv
\sin\vartheta^{}_x$, and so on. One can see that the CP-violating
phase $\delta^{}_{\rm q}$ just appears in the four corners of $V$.
Confronting Eq. (20) with current experimental data leads us to
$\vartheta^{}_x \simeq 13.2^\circ$, $\vartheta^{}_y \simeq
10.1^\circ$, $\vartheta^{}_z \simeq 10.3^\circ$ and $\delta^{}_{\rm
q} \simeq 1.1^\circ$ \cite{GX}, consistent with $\alpha \simeq
90^\circ$ or equivalently ${\rm Re} (V^{}_{tb} V^{}_{ud} V^*_{td}
V^*_{ub}) \simeq 0$. Note that $\vartheta^{}_x$ and $\delta^{}_{\rm
q}$ are essentially stable when the energy scale changes, but
$\vartheta^{}_y$ and $\vartheta^{}_z$ may slightly be modified due
to radiative corrections. To see this point more clearly, let us
take account of the approximate one-loop renormalization-group
equations (RGEs) of nine CKM matrix elements \cite{Babu}:
\begin{eqnarray}
\frac{\rm d}{{\rm d} t} \ln |V^{}_{ud}| & \simeq & \frac{\rm d}{{\rm
d} t} \ln |V^{}_{cs}| \simeq \frac{\rm d}{{\rm d} t} \ln |V^{}_{tb}|
\simeq \frac{\rm d}{{\rm d} t} \ln |V^{}_{us}| \simeq \frac{\rm
d}{{\rm d} t} \ln |V^{}_{cd}| \simeq 0 \; ,
\nonumber \\
\frac{\rm d}{{\rm d} t} \ln |V^{}_{ub}| & \simeq & \frac{\rm d}{{\rm
d} t} \ln |V^{}_{cb}| \simeq \frac{\rm d}{{\rm d} t} \ln |V^{}_{td}|
\simeq \frac{\rm d}{{\rm d} t} \ln |V^{}_{ts}| \simeq c \left( y^2_t
+ y^2_b \right) \; ,
\end{eqnarray}
where $t \equiv (1/16\pi^2)\ln (\mu/M^{}_Z)$ for a given energy
scale $\mu$ above the electroweak scale, $y^{}_t$ and $y^{}_b$
standard for the Yukawa coupling eigenvalues of top and bottom
quarks, $c = -3$ (or $-1$) holds in the SM (or its supersymmetric
extension). Combining Eq. (20) with Eq. (21), we obtain the
approximate one-loop RGEs of $\vartheta^{}_x$, $\vartheta^{}_y$,
$\vartheta^{}_z$ and $\delta^{}_{\rm q}$ as follows:
\begin{eqnarray}
\frac{\rm d}{{\rm d} t} \ln \sin\vartheta^{}_x & \simeq & \frac{\rm
d}{{\rm d} t} \ln \sin\delta^{}_{\rm q} \simeq 0 \; ,
\nonumber \\
\frac{\rm d}{{\rm d} t} \ln \sin\vartheta^{}_y & \simeq & \frac{\rm
d}{{\rm d} t} \ln \sin\vartheta^{}_z \simeq c \left( y^2_t + y^2_b
\right) \; .
\end{eqnarray}
Hence the mixing angle $\vartheta^{}_x$ and the CP-violating phase
$\delta^{}_{\rm q}$ are almost stable against the RGE running
effects, and the mixing angles $\vartheta^{}_y$ and $\vartheta^{}_z$
are expected to have the same RGE running behaviors at the one-loop
level. Because of ${\cal J}^{}_{\rm q} = c^{}_x s^2_x c^{}_y s^{}_y
c^{}_z s^{}_z \sin\delta^{}_{\rm q}$, the RGE evolution of ${\cal
J}^{}_{\rm q}$ is mainly controlled by that of $s^{}_y$ and
$s^{}_z$.

\section{A brief comment on the strong CP problem}

So far we have only paid attention to weak CP violation based on the
CKM matrix in the SM. Here let us make a brief comment on the strong
CP problem, because it is closely related to the quark masses and
may naturally disappear if one of the six quark masses vanishes. It
is well known that there exists a P- and T-violating term ${\cal
L}^{}_\theta$, which comes from the instanton solution to the $\rm
U(1)^{}_{\rm A}$ problem \cite{Weinberg}, in the Lagrangian of QCD
for strong interactions of quarks and gluons \cite{QCD}. This
CP-violating term can be compared with the mass term of six quarks,
${\cal L}^{}_{\rm m}$, as follows:
\begin{eqnarray}
{\cal L}^{}_\theta & = & \theta \frac{\alpha^{}_{\rm s}}{8\pi} \
G^a_{\mu \nu} \tilde{G}^{a\mu \nu} \; ,
\nonumber \\
{\cal L}^{}_{\rm m} & = & \overline{\left( \matrix{u & c & t & d & s
& b}\right)^{}_{\rm L}} \ {\cal M} \left(\matrix{ u \cr c \cr t \cr
d \cr s \cr b}\right)^{}_{\rm R} + {\rm h.c.} \; ,
\end{eqnarray}
where $\theta$ is a free dimensionless parameter characterizing the
presence of CP violation, $\alpha^{}_{\rm
s}$ is the strong fine-structure constant, $G^a_{\mu \nu}$ (for
$a=1,2,\cdots,8$) denote the $\rm SU(3)^{}_{\rm c}$ gauge fields,
$\tilde{G}^{a\mu \nu} \equiv
\epsilon^{\mu\nu\alpha\beta}G^a_{\mu\nu}/2$, and $\cal M$ is the
overall $6\times 6$ quark mass matrix. The chiral transformation of
the quark fields $q\to \exp(i\phi^{}_q \gamma^{}_5) q$ (for $q = u,
c, t; d, s, b$) leads to the changes
\begin{eqnarray}
&& \theta \to \theta - 2 \sum_q \phi^{}_q \; ,
\nonumber \\
&& \arg\left(\det {\cal M}\right) \to \arg\left(\det {\cal M}\right)
+ 2 \sum_q \phi^{}_q \; ,
\end{eqnarray}
in which the change of $\theta$ follows from the chiral anomaly
\cite{Anomaly} in the chiral currents
\begin{eqnarray}
\partial^{}_\mu \left(\overline{q} \gamma^\mu \gamma^{}_5 q \right)
= 2i m^{}_q \overline{q} \gamma^{}_5 q + \frac{\alpha^{}_{\rm
s}}{4\pi} G^a_{\mu\nu} \tilde{G}^{a \mu\nu} \; .
\end{eqnarray}
Then the effective CP-violating term in QCD, which is invariant under
the above chiral transformation, turns out to be
\begin{eqnarray}
{\cal L}^{}_{\overline\theta} = \overline{\theta}
\frac{\alpha^{}_{\rm s}}{8\pi} \ G^a_{\mu \nu} \tilde{G}^{a\mu \nu} \; ,
\end{eqnarray}
where $\overline{\theta} = \theta + \arg\left(\det {\cal M}\right)$
is a sum of the QCD and electroweak contributions \cite{Review}. The
latter depends on the phase structure of the quark mass matrix $\cal
M$. Because of
\begin{eqnarray}
\left|\det {\cal M}\right| = m^{}_u m^{}_c m^{}_t m^{}_d m^{}_s
m^{}_b \; ,
\end{eqnarray}
the determinant of $\cal M$ becomes vanishing in the $m^{}_u \to 0$
(or $m^{}_d \to 0$) limit. In this case the phase of $\det {\cal M}$
is arbitrary, and thus it can be arranged to cancel out $\theta$
such that $\overline{\theta} \to 0$. Namely, QCD would be a
CP-conserving theory if one of the six quarks were massless. But
current experimental data have definitely ruled out the possibility
of $m^{}_u =0$ or $m^{}_d =0$. Moreover, the experimental upper
limit on the neutron electric dipole moment yields
$\overline{\theta} < 10^{-10}$ \cite{EDM}. The strong CP problem is
therefore a theoretical problem of how to explain why
$\overline{\theta}$ is nonzero but so small \cite{Peccei}.

A comparison between the weak and strong CP-violating effects might
make sense, but it is difficult to choose a proper measure for
either of them. The issue involves the reference energy scale and
flavor parameters which may directly or indirectly determine the
strength of CP violation. To illustrate, we consider the following
preliminary measures of weak and strong CP-violating effects in the
SM
\footnote{We admit that running the heavy quark masses $m^{}_c$,
$m^{}_b$ and $m^{}_t$ down to the QCD scale might not make sense
\cite{Koide}. One may only consider the masses of up and down quarks
\cite{Huang} and then propose ${\rm CP^{}_{strong}} \sim m^{}_u
m^{}_d \sin\overline{\theta}/\Lambda^2_{\rm QCD}$ as an alternative
measure of strong CP violation.}:
\begin{eqnarray}
&& {\rm CP^{}_{\rm weak}} \sim \frac{1}{\Lambda^6_{\rm EW}}
\left(m^{}_u - m^{}_c\right) \left(m^{}_c - m^{}_t\right)
\left(m^{}_t - m^{}_u\right) \left(m^{}_d - m^{}_s\right)
\left(m^{}_s - m^{}_b\right) \left(m^{}_b - m^{}_d\right) {\cal
J}^{}_{\rm q} \sim 10^{-13} \; , \nonumber \\
&& {\rm CP^{}_{\rm strong}} \sim \frac{1}{\Lambda^6_{\rm QCD}}
m^{}_u m^{}_c m^{}_t m^{}_d m^{}_s m^{}_b \sin\overline{\theta} \sim
10^{4} \sin\overline{\theta} < 10^{-6} \; ,
\end{eqnarray}
where $\Lambda^{}_{\rm EW} \sim 10^2$ GeV, $\Lambda^{}_{\rm QCD}
\sim 0.2$ GeV, and the sine function of $\overline{\theta}$ has been
adopted to take account of the periodicity in its values. So
the effect of weak CP violation would vanish if the masses of any two
quarks in the same (up or down) sector were equal
\footnote{In this special case one of the three mixing angles of $V$
must vanish, leading to ${\cal J}^{}_{\rm q} =0$ too \cite{FX99}.},
and the effect of strong CP violation would vanish if $m^{}_u \to 0$
or $\sin\overline{\theta} \to 0$ held. The significant suppression
of CP violation in the SM implies that an interpretation of the
observed matter-antimatter asymmetry of the Universe \cite{PDG}
requires a new source of CP violation beyond the SM, such as
leptonic CP violation in the decays of heavy Majorana neutrino based
on the seesaw and leptogenesis mechanisms \cite{FY} in neutrino
physics.

\section{Summary}

We have pointed out that it is possible to partly understand the observed
pattern of quark flavor mixing in the chiral and heavy quark mass
limits. Such a model-independent access to the underlying quark
flavor structure can help us explain $|V^{}_{us}| \simeq
|V^{}_{cd}|$, $|V^{}_{cb}| \simeq |V^{}_{ts}|$,
$|V^{}_{cd}/V^{}_{td}| \simeq |V^{}_{cs}/V^{}_{ts}| \simeq
|V^{}_{tb}/V^{}_{cb}|$, and $|V^{}_{ub}/V^{}_{cb}| <
|V^{}_{td}/V^{}_{ts}|$. In particular, we have argued that the
phenomenologically successful relations $|V^{}_{ub}/V^{}_{cb}| \sim
\sqrt{m^{}_u/m^{}_c}$ and $|V^{}_{td}/V^{}_{ts}| \simeq
\sqrt{m^{}_d/m^{}_s}$ can be reasonably conjectured in the heavy
quark mass limits. In view of the experimental fact that two of the
CKM unitarity triangles are almost the right triangles with $\alpha
\simeq 90^\circ$, we have obtained ${\cal J}^{}_{\rm q} \simeq
|V^{}_{ud}| \cdot |V^{}_{ub}| \cdot |V^{}_{td}| \cdot |V^{}_{tb}|$.
A particular parametrization of $V$ with the minimal CP-violating phase
has been emphasized, and the RGE running behaviors of its parameters
have been discussed. We have also made a very brief comment on the strong
CP problem, and compared between the preliminary measures of strong
and weak CP-violating effects in the quark sector within the SM.

Although our present attempts in this regard remain quite limited,
we do have obtained some encouraging results. We hope that the
underlying flavor theory, which might be related to a certain flavor
symmetry and its spontaneous or explicit breaking mechanism, could
provide us with a more convincing dynamical reason for what we have
observed about the structure of quark flavor mixing and CP violation.

One may naturally ask whether the leptonic flavor mixing structure
could similarly be understood in the reasonable mass limits of the
charged leptons and neutrinos. While the charged leptons have a strong
mass hierarchy, the neutrino mass spectrum remains unknown to us
--- we do not know whether the three neutrinos have a normal mass
hierarchy $m^{}_1 < m^{}_2 < m^{}_3$ or an inverted mass hierarchy
$m^{}_3 < m^{}_1 < m^{}_2$. On the other hand, our knowledge on the
MNSP matrix is still poor because its CP-violating phases are all
undetermined. Hence a lot of more experimental and theoretical efforts
are needed to make towards a better understanding of the flavor issues
in the lepton sector.

\vspace{0.3cm}

The author is deeply indebted to Shun Zhou for helpful comments in
Oberw$\rm\ddot{o}$lz and useful discussions via emails. This work
was supported in part by the Ministry of Science and Technology of
China under grant No. 2009CB825207, and in part by the National
Natural Science Foundation of China under grant No. 11135009.

\newpage

\appendix
\section{}

Given the $3\times 3$ quark mass matrices $M^{}_{\rm u}$ and
$M^{}_{\rm d}$ in an arbitrary flavor basis, it is always possible to
make an appropriate basis transformation such that the resulting mass
matrices $\overline{M}^{}_{\rm u}$ and $\overline{M}^{}_{\rm d}$
simultaneously have vanishing (1,1), (2,2), (1,3) and (3,1) elements
\cite{Branco}:
\begin{eqnarray}
\overline{M}^{}_{\rm u} = \left( \matrix{0 & X^{}_{\rm u} & 0 \cr
X^\prime_{\rm u} & 0 & Y^{}_{\rm u} \cr 0 & Y^\prime_{\rm u} &
Z^{}_{\rm u} \cr} \right) \; ,
\nonumber \\
\overline{M}^{}_{\rm d} = \left( \matrix{0 & X^{}_{\rm d} & 0 \cr
X^\prime_{\rm d} & 0 & Y^{}_{\rm d} \cr 0 & Y^\prime_{\rm d} &
Z^{}_{\rm d} \cr} \right) \; .
\end{eqnarray}
The determinants of $\overline{M}^{}_{\rm u}$
and $\overline{M}^{}_{\rm d}$ turn out to be
\begin{eqnarray}
\left|\det\overline{M}^{}_{\rm u}\right| & = & \left|X^{}_{\rm u}
X^\prime_{\rm u} Z^{}_{\rm u}\right| = m^{}_u m^{}_c m^{}_t \; ,
\nonumber \\
\left|\det\overline{M}^{}_{\rm d}\right| & = & \left|X^{}_{\rm d}
X^\prime_{\rm d} Z^{}_{\rm d}\right| = m^{}_d m^{}_s m^{}_b \; .
\end{eqnarray}
In this basis one may simply set $X^{}_{\rm u} \to 0$ (or $X^{}_{\rm
d} \to 0$) to achieve the chiral quark mass limit $m^{}_u \to 0$ (or
$m^{}_d \to  0$), or vice versa. Defining $\overline{H}^{}_{\rm q}
\equiv \overline{M}^{}_{\rm q} \overline{M}^\dagger_{\rm q}$ (for
$\rm q = u$ or $\rm d$), we then obtain
\begin{eqnarray}
\lim_{m^{}_u \to 0} \overline{H}^{}_{\rm u} & = & \left( \matrix{0 &
0 & 0 \cr 0 & |X^\prime_{\rm u}|^2 + |Y^{}_{\rm u}|^2 & Y^{}_{\rm u}
Z^*_{\rm u} \cr 0 & Y^*_{\rm u} Z^{}_{\rm u} & |Y^\prime_{\rm u}|^2
+ |Z^{}_{\rm u}|^2 \cr} \right) \; ,
\nonumber \\
\lim_{m^{}_d \to 0} \overline{H}^{}_{\rm d} & = & \left( \matrix{0 &
0 & 0 \cr 0 & |X^\prime_{\rm d}|^2 + |Y^{}_{\rm d}|^2 & Y^{}_{\rm d}
Z^*_{\rm d} \cr 0 & Y^*_{\rm d} Z^{}_{\rm d} & |Y^\prime_{\rm d}|^2
+ |Z^{}_{\rm d}|^2 \cr} \right) \; .
\end{eqnarray}
Hence Eq. (5) is the result of a specific basis choice instead of a
pure assumption. But we admit that a given quark mass limit does not
uniquely correspond to a definite texture of the quark mass matrix,
simply because the latter is basis-dependent.

To illustrate the above point in a more transparent way, let us take
a look at the following typical example of quark mass matrices in
two different bases
\cite{FX00}:
\begin{eqnarray}
M^{(\rm H)}_{\rm q} & = & A^{}_{\rm q} \left(\matrix{0 & 0 & 0 \cr 0
& 0 & 0 \cr 0 & 0 & 1 \cr} \right) \; ,
\nonumber \\
M^{(\rm D)}_{\rm q} & = & \frac{A^{}_{\rm q}}{3} \hspace{-0.05cm}
\left(\matrix{1 & 1 & 1 \cr 1 & 1 & 1 \cr 1 & 1 & 1 \cr} \right) \;
,
\end{eqnarray}
where $\rm q = u$ or $\rm d$. It is well known that the democratic
texture $M^{(\rm D)}_{\rm q}$ can be transformed into the
hierarchical texture $M^{(\rm H)}_{\rm q}$ via $U^{}_0 M^{(\rm
D)}_{\rm q} U^\dagger_0 = M^{(\rm H)}_{\rm q}$, where
\begin{eqnarray}
U^{}_0 = \left(\matrix{\frac{1}{\sqrt 2} & \frac{-1}{\sqrt 2} & 0
\cr \frac{1}{\sqrt 6} & \frac{1}{\sqrt 6} & \frac{-2}{\sqrt 6} \cr
\frac{1}{\sqrt 3} & \frac{1}{\sqrt 3} & \frac{1}{\sqrt 3} \cr}
\right) \; ,
\end{eqnarray}
which is actually the leading term of the democratic flavor mixing
pattern \cite{FX96}. We may obtain $m^{}_u = m^{}_c =0$ (or $m^{}_d
= m^{}_s =0$) from either $M^{(\rm D)}_{\rm u}$ (or $M^{(\rm
D)}_{\rm d}$) or $M^{(\rm H)}_{\rm u}$ (or $M^{(\rm H)}_{\rm d}$),
but their textures are apparently different. If a diagonal
perturbation of the form $\Delta M^{}_{\rm q} \propto {\rm Diag}\{0,
0, A^\prime_{\rm q}\}$ (for $|A^\prime_{\rm q}| \ll |A^{}_{\rm q}|$)
is simultaneously added to $M^{(\rm D)}_{\rm q}$ and $M^{(\rm
H)}_{\rm q}$, one will arrive at $m^{}_u = m^{}_d =0$ without any
nontrivial quark flavor mixing in the hierarchical case, but $m^{}_u
= m^{}_d =0$ with a nontrivial flavor mixing effect between the
second and third quark families in the democratic case.
This observation clearly
illustrates the point that a specific quark mass limit may
correspond to quite different forms of the quark mass matrix in
different flavor bases, leading to quite different flavor mixing
effects.

\begin{figure*}
\vspace{1cm} \centering
\includegraphics[width=7cm]{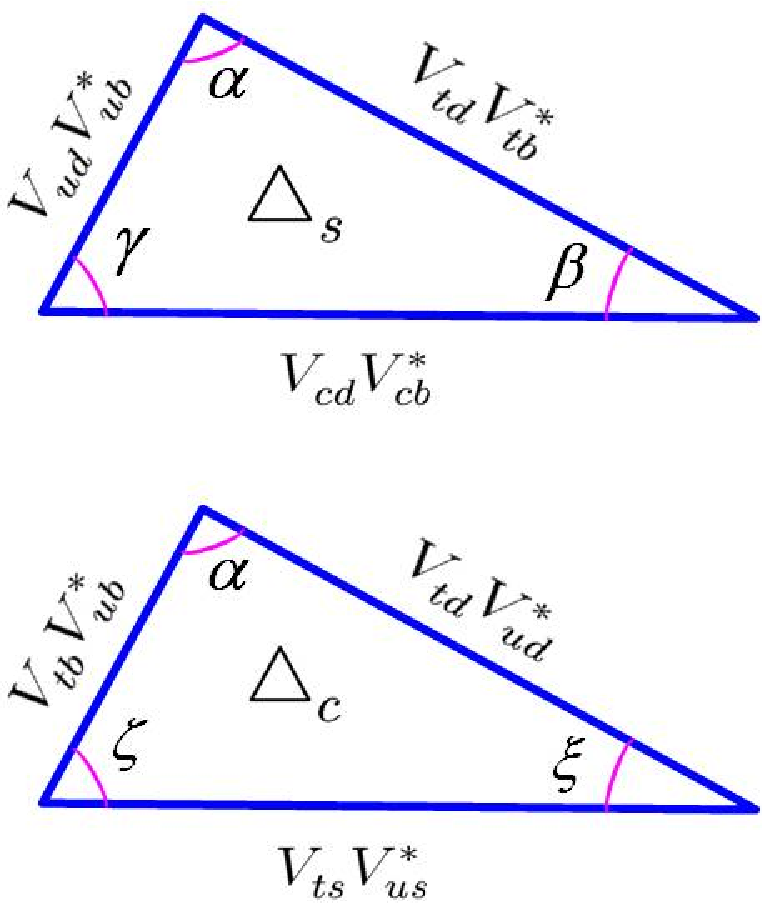}
\vspace{0.6cm} \caption{The CKM unitarity triangles $\triangle^{}_s$
and $\triangle^{}_c$ in the complex plane. They share a common
inner angle $\alpha$, which is essentially equal to $90^\circ$
as indicated by current experimental data.}
\end{figure*}
\begin{figure*}
\vspace{0.8cm} \centering
\includegraphics[width=6.5cm]{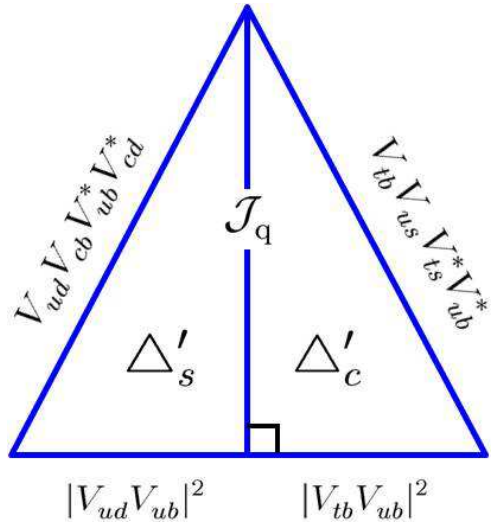}
\vspace{0.6cm} \caption{The rescaled CKM unitarity triangles
$\triangle^\prime_s$ and $\triangle^\prime_c$, which share a common
side equal to the Jarlskog invariant ${\cal J}^{}_{\rm q}$ (due to
$\alpha = 90^\circ$), in the complex plane.}
\end{figure*}
\end{document}